# Magnetic Exciton-Polariton with Strongly Coupled Atomic and Photonic Anisotropies

Qiuyang Li,[1,5] Xin Xie,[1,5] Adam Alfrey,[2] Christiano W. Beach,[1] Nicholas McLellan,[1] Yang Lu,[3] Jiaqi Hu,[2] Wenhao Liu,[4] Nikhil Dhale[4], Bing Lv,[4] Liuyan Zhao,[1] Kai Sun,[1] Hui Deng[1,2,3],*

[1]Department of Physics, University of Michigan, Ann Arbor, Michigan 48109, United States
[2]Applied Physics Program, University of Michigan, Ann Arbor, Michigan 48109, United States
[3]Department of Electrical Engineering and Computer Science, University of Michigan, Ann Arbor, Michigan 48109, United States
[4]Department of Physics, University of Texas at Dallas, Richardson, Texas 75080, United States
[5]These authors contributed equally to this work.
*Corresponding author. Email: dengh@umich.edu

**Anisotropy plays a key role in science and engineering. However, the interplay between the material and engineered photonic anisotropies has hardly been explored due to the vastly different length scales. Here we demonstrate a matter-light hybrid system, exciton-polaritons in a 2D antiferromagnet, CrSBr, coupled with an anisotropic photonic crystal (PC) cavity, where the spin, atomic lattice, and photonic lattices anisotropies are strongly correlated, giving rise to unusual properties of the hybrid system and new possibilities of tuning. We show exceptionally strong coupling between engineered anisotropic optical modes and anisotropic excitons in CrSBr, which is stable against excitation densities a few orders of magnitude higher than polaritons in isotropic materials. Moreover, the polaritons feature a highly anisotropic polarization tunable by tens of degrees by controlling the matter-light coupling via, for instance, spatial alignment between the material and photonic lattices, magnetic field, temperature, cavity detuning and cavity quality-factors. The demonstrated system provides a prototype where atomic- and photonic-scale orders strongly couple, opening opportunities of photonic engineering of quantum materials and novel photonic devices, such as compact, on-chip polarized light source and polariton laser.**

Anisotropy is a fundamental property of materials widely exploited in science and engineering. While often arising from microscopic structures of the materials on the Angstrom scale, such as lattice symmetries of crystals,[1] spin textures of magnets,[2,3] and orientations of molecular bonds,[4] it manifests in many elementary macroscopic properties, including magnetism,[2,3] anisotropic electrical conduction,[5] and polarized optical responses.[6,7] Tuning the anisotropy, however, often requires substantial change to the materials and is thus limited. Inspired by natural anisotropy, anisotropic properties of light have also been engineered via artificial structures on the wavelength scale, such as photonic crystals and metamaterials.[8,9] Interplay between the material and engineered photonic anisotropies, however, has hardly been explored due to the vastly different length scales.

Here, we demonstrate a matter-light hybrid system, exciton-polaritons in a 2D magnet, CrSBr, coupled with an anisotropic photonic crystal (PC) cavity, where the spin, atomic lattice, and photonic lattice anisotropies become correlated, giving rise to unusual properties of the hybrid system and new possibilities of tuning. We show exceptionally strong coupling between engineered anisotropic optical modes and anisotropic excitons in CrSBr, when at the presence of atomic-structure and spin anisotropies in the 2D magnet. The resulting new quasi-particles, polaritons, are stable against excitation densities a few orders of magnitude higher than polaritons in isotropic materials, as well as against increasing temperature up to the Néel temperature of the 2D magnet. However, they disappear when the magnetism is destroyed. Due to the strong coupling between the atomic and photonic lattice anisotropies, the polaritons feature a highly anisotropic polarization tunable by 10s of degrees by many means, such as cavity resonance quality factor and detuning, magnetic field and temperature that tunes the exciton resonance, and spatial alignment between the material and photonic lattices that directly tunes the coupling. Such a polariton system may enable novel polarization photonics[10,11] and further studies of coherent interactions among spin, charge and light.[12-15]

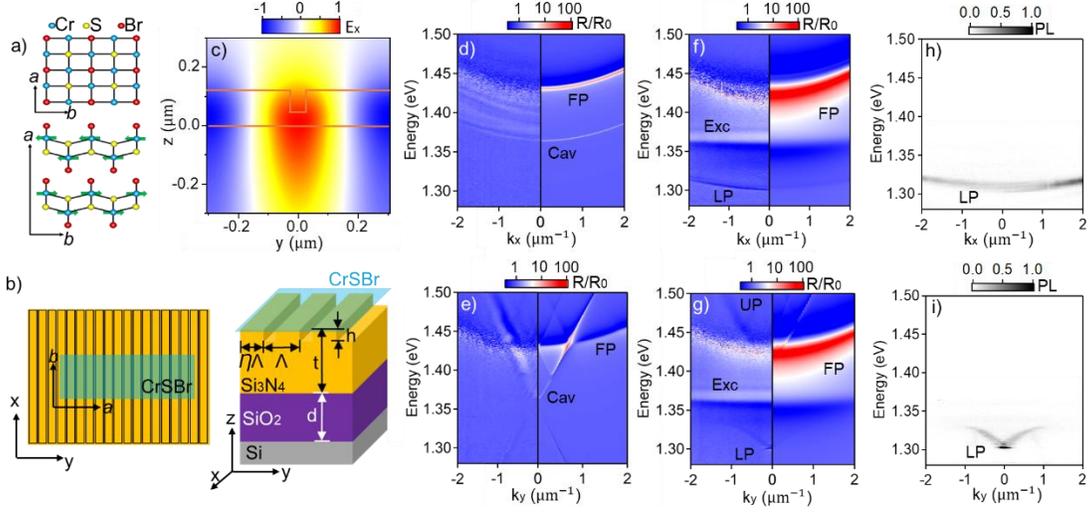

**Fig. 1. Strongly coupled exciton-polariton in aligned CrSBr-PC.** (a) Schematic of CrSBr lattice. Green arrows mark the magnetization orientation on each Cr atom in different layers. (b) Schematics of a 7L-CrSBr integrated with a PC for polariton studies. (c) PC cavity mode profile shown as distribution of the x-component of the electric field amplitude ($E_x$) of mode with zero in-plane wavenumber. The orange solid lines mark the boundary of the $Si_3N_4$ grating in one period. (d-g) Momentum-resolved, normalized reflection spectra $R/R_0$ of (d, e) the bare PC and (f, g) CrSBr-PC dispersed along (d, f) x- and (e, g) y-direction. R is the reflection from the device and $R_0$ is the reflection from the flat $Si_3N_4$ surfaces. The left panels of (d-g) are experimental results; the right panels are simulation results. The FP mode results from multiple reflection at the interfaces between air, $Si_3N_4$, $SiO_2$, and Si substrate. (h-i) Momentum-resolved PL spectrum of CrSBr-PC dispersed along (h) $k_x$- and (i) $k_y$-direction.

To create the anisotropic, magnetic polariton system, we integrate an atomically thin 7-layer (7L) CrSBr crystal with a 113 nm-thick slab photonic crystal (PC) as a cavity. The material, CrSBr, is a special 2D magnetic semiconductor, where its lattice and magnetic anisotropies lead to tightly bound excitons with both an exceptionally large oscillator strength and strong linear polarization. CrSBr has an orthorhombic lattice structure (Fig. 1a). It features 2D antiferromagnetism (AFM), where each CrSBr monolayer is a ferromagnet (FM) with a spontaneous magnetization along the b-axis but in opposite directions between adjacent layers.[16,17] Due to the atomic lattice and magnetic anisotropies, CrSBr feature quasi-1D excitons,[18] that are linearly polarized along the b-axis,[16,18] with tight exciton binding and large oscillator strength. Energy of the excitons is tunable by both magnons and static magnetic fields[12,13,16]. These properties make CrSBr uniquely suited for creating tunable, anisotropic, and magnetic polaritons.

We chose a 7L-CrSBr to ensure 2D magnetism in the few-layer limit, uniform and large per-molecular layer exciton-photon coupling (the thinner the better), and at the same large total Rabi-splitting (the thicker the better). Before placing it onto the PC, we identify the exciton resonance in our 7L-CrSBr at 1.361±0.002 eV by reflection contrast (RC) and photoluminescence (PL) measurements (Supplementary Fig. 1a). Polarization-resolved RC and PL spectra confirm that the exciton is linear polarized along the b-axis (Supplementary Figs. 1b-d), consistent with previous results.[16,18]

To form CrSBr exciton-polaritons, the 7L CrSBr crystal is placed on an anisotropic 1D PC cavity, consisting of a partially etched $Si_3N_4$ grating on $SiO_2$ substrate (Fig. 1b).[19,20] Details of the PC and its fabrication are in Methods. The cavity mode is transverse-electric (TE) polarized along the grating bar, the x-direction. Figs. 1d and 1e show the momentum-resolved normalized reflection spectra ($R/R_0$, where R is reflection from the device, normalized by $R_0$, reflection from the flat $Si_3N_4$ surfaces) of the bare cavity, without CrSBr, along the x- and y-direction, respectively. We observe a parabolically-dispersed cavity mode along the x-direction (fitted with gray dashed line) with a minimum of 1.364±0.002 eV at an in-plane wavenumber $k_{\parallel}$~0. The quality factor (Q-factor), which is estimated from its full width at the half maximum (FWHM, 0.6±0.1 meV), is ~1000. Along the y-direction, a much steeper dispersion is observed as expected. The simulated mode profile of the cavity mode (Fig. 1c, and see Methods for simulation details) confirms a large field amplitude at the surface of $Si_3N_4$ for enhanced coupling with a thin 2D material placed on top.

Anisotropy of the integrated CrSBr-PC system is studied by changing the alignment between the atomic and photonic lattices and characterizing the modes via momentum-resolved normalized reflection and PL spectroscopies.

We first study the impact of atomic lattice anisotropy alone on the exciton-photon coupling by aligning it to the photonic lattice anisotropy, i.e., aligning the b-axis of CrSBr with the x-axis of the PC, as illustrated in Fig. 1b. In this case, the resulting mode should follow the same anisotropy as CrSBr and PC, and the light-matter coupling is also maximized.

Energy-momentum dispersions of the resonances in the integrated system are measured along both the x-direction (Figs. 1f&h) and y-direction (Figs. 1g&i) in both normalized reflection and PL. A dispersive lower polariton (LP) band, with the same linear polarization as the exciton, is observed in all spectra, well below the uncoupled exciton or cavity modes, which evidences strong coupling. Along the y-direction, the steeper cavity dispersion makes the upper polariton (UP) band also observable in the normalized reflection spectrum (Fig. 1g), distinguished from the isotropic Fabry-Pérot (FP) mode of the $Si_3N_4$ layer near ~1.42 eV.[20] In normalized reflection spectra (Figs. 1f&g), we also observe the non-dispersive exciton band (marked with a flat gray dashed line) due to coupling of the exciton with free-space modes. All the cavity, exciton, and polariton bands are reproduced by the rigorous coupled wave analysis (RCWA) simulation (right panels of Figs. 1d-g) with excellent agreement. The LP bands are also clearly observed in PL spectra (Figs. 1h&i). The exciton and UP bands are largely absent in PL spectra due to energy relaxation of the particles to the LP band .

The measured dispersions $E_{LP,UP}(k)$ can be compared with polariton dispersions derived from the two coupled oscillator model. Modeling the polaritons as eigen-modes of strongly coupled exciton and cavity photon modes under the rotating wave approximation, we have:

$$E_{LP,UP}(k) = \frac{1}{2}\left[E_{exc} + E_{cav}(k) + i\frac{(w_{cav} + w_{exc})}{4}\right]$$
$$\pm \sqrt{g_0^2 + \frac{[E_{exc} - E_{cav}(k) + i(w_{cav} - w_{exc})/2]^2}{4}} \quad (1)$$

$$2\hbar\Omega = \sqrt{4g_0^2 - (w_{cav} - w_{exc})^2/4} \quad (2)$$

where $E_{exc}$ and $E_{cav}$ are the energies of the uncoupled exciton and cavity mode, respectively, and $w_{exc}$ and $w_{cav}$ are their FWHM; $g_0$ is the coupling strength and $2\hbar\Omega$ is the vacuum Rabi splitting. We obtain $E_{cav}(k)$ independently by fitting the bare cavity dispersion to a parabola for $k_x$ direction (Supplementary Fig. 2a, grey dashed line) and two linear lines for $k_y$ direction (Supplementary Fig. 2c, grey dashed line). Note that $E_{cav}(k)$ does not shift with or without CrSBr on top, which is confirmed with another CrSBr-PC sample where the cavity mode is far detuned from the resonance of CrSBr (Supplementary Fig. 3). We also directly measure $E_{exc}$=1.361±0.002 eV and $w_{exc}$=10±1 meV from RC and PL of 7L-CrSBr on a flat $Si_3N_4$ (Supplementary Fig. 1a). Using these results, we perform a global fit of the measured polariton dispersions along $k_x$ and $k_y$ using Eq. (1), with the coupling strength $g_0$ as the only fitting parameter. The measured dispersions fit very well (Supplementary Fig. 2). From the fitting, we obtain $g_0 = 58 \pm 2$ meV and $2\hbar\Omega = 116 \pm 3$ meV. These values are much larger than the exciton and cavity linewidths. The well fitted polariton dispersions and large coupling strength confirm that the system is deep in the strong-coupling regime.

The measured vacuum Rabi splitting per molecular layer is $2\hbar\Omega/\sqrt{N} = 44 \pm 1$ meV for $N = 7$. It is more than double the highest value reported for transition metal dichalcogenides ($MX_2$, M=Mo, W; X=Se, S) monolayers on a similar PC[20,21] or in a DBR-based cavity,[22] and is over an order of magnitude larger than another 2D magnet, $NiPS_3$, in a DBR-based cavity.[23] Such a large Rabi splitting suggests a large oscillator strength due to the quasi-1D character of the excitons in the anisotropic crystal lattice. Note that the recently reported strong coupling of bulk CrSBr crystals in large DBR-based cavities[14,15] show a larger total coupling strength, due to the much thicker crystal (>400 nm in thickness), but a much smaller per layer oscillator strength, due to highly non-uniform cavity field distribution across the bulk crystal.

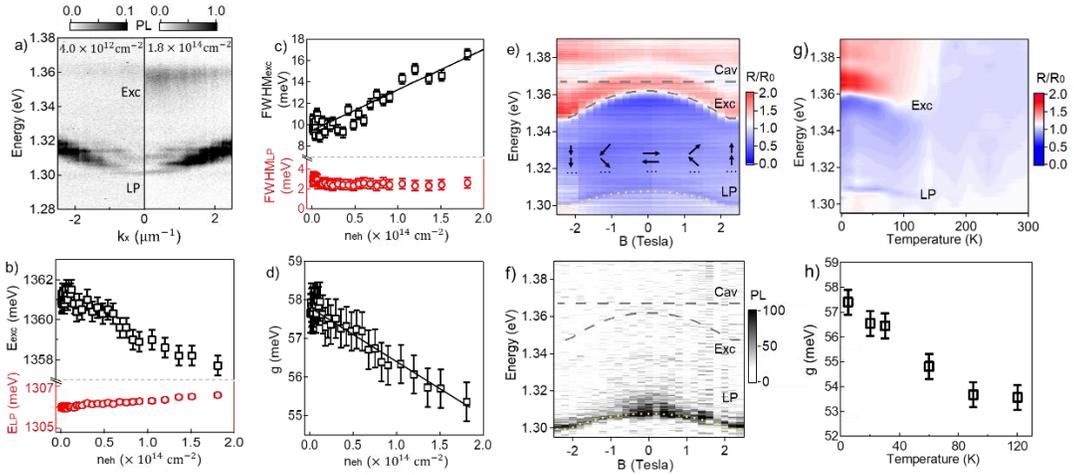

**Fig. 2. Excitation density, external magnetic field, and temperature dependence of CrSBr-PC polaritons.** (a) Momentum-resolved PL spectra at e-h density of $4.0\times10^{12}$ cm$^{-2}$ (left panel) and $1.8\times10^{14}$ cm$^{-2}$ (right panel). (b) Exciton energy (black squares) and LP energy (red circles) as a function of e-h density ($n_{eh}$). (c) FWHM of exciton (black squares) and LP (red circles) as a function of e-h density. The black solid line is a linear fit for the FWHM of exciton. (d) Coupling strength g as a function of e-h density. The black solid line is a linear fit. (e) Normalized reflection and (f) PL spectra of aligned CrSBr-PC at $k_x=0$

as a function of magnetic field B. The gray dashed lines mark the uncoupled exciton and cavity modes. The yellow dotted lines are fits of the LP bands. Inset of (e) is the schematic of magnetic order at different external magnetic fields, where the black arrows mark the magnetization (spin orientation) for each CrSBr layer. (g) Reflection spectra of aligned CrSBr-PC at $k_x=0$ as a function of temperature. (h) Coupling strength g as a function of temperature.

To further study the effects of the quasi-1D character of the excitons, we measure the excitation density dependence of the excitons and polaritons. Fig. 2a shows the comparison of momentum-resolved PL spectra of CrSBr-PC at low ($n_{eh} = 4 \times 10^{12}$ cm$^{-2}$, left panel) and high ($n_{eh} = 1.8 \times 10^{14}$ cm$^{-2}$, right panel) e-h density as two examples. From the spectra at different $n_{eh}$, we obtain the e-h density dependence of exciton and LP energy (Fig. 2b) and FWHM (Fig. 2c), and the exciton-cavity coupling strength, g (Fig. 2d). These results show an unusually robust polariton mode compared to isotropic 2D semiconductors.

As shown in Fig. 2b, the exciton resonance has a small redshift at high e-h densities, which results from a combination of bandgap renormalization, screening and phase space filling.[24,25] The exciton FWHM broadens linearly with increasing $n_{eh}$ while the FWHM of LP barely changes with $n_{eh}$ (Fig. 2c).

The LP shows only a small blueshift (Fig. 2b), it remains clearly resolved and far separated from the exciton resonance or the cavity resonance above the exciton, up to $n_{max}=(1.8\pm0.1)\times10^{14}$ cm$^{-2}$. This observation of the robust polariton mode is corroborated by the very slight reduction of the coupling strength by less than 5% up to $n_{max}$ (Fig. 2d). The approximately linear dependence of g with $n_{eh}$ suggests the system remains well below the exciton saturation or the Mott transition.[26] Fitting the data with $g(n_{eh}) = g_0(1 - \frac{n_{eh}}{7n_s})$, we obtain an exciton saturation density per layer, $n_s=(5.4\pm0.2)\times10^{14}$ cm$^{-2}$. This value is over one order of magnitude larger than typical values of transitional metal dichalcogenide monolayers,[27] and three orders of magnitude larger than those of III-As.[26] From the $n_s$, we estimate the effective exciton Bohr radius ($a_B$) to be a mere 2.5$\pm$0.1 Å for a round-shaped exciton, ~3-fold smaller than the one of a similar 2D magnet, NiPS$_3$.[23] Such a tightly bound exciton has been suggested, as a result of the strong lattice anisotropy and quasi-1D character of excitons in CrSBr.[18] Considering anisotropic dielectric constants of 43.1 and 11.5 along the b- and a-axis, respectively,[14] the Bohr radius is estimated to be 4.8$\pm$0.1 and 1.3$\pm$0.1 Å for an oval shaped exciton. The small Bohr radius is consistent with the measured very large coupling strength and the calculated exciton binding energy of 0.5-0.9 eV.[16,18]

Unique to CrSBr excitons and polaritons, their properties are strongly correlated with not only the lattice anisotropy, but also the spin anisotropies in the magnetically ordered phase. These spin anisotropies are controllable by an external magnetic field (B) and temperature. Figs. 2e&f show the normalized reflection and PL spectra of the CrSBr-PC system at $k_x=0$ as a function of an out-of-plane B field. The exciton energy redshifts from $1.361\pm0.002$ meV at B = 0 T to $1.347\pm0.002$ meV at $|B| >2$ T, similar to previous reports.[16] It has been understood as resulting from increased coupling between adjacent CrSBr layers as their spins are rotated from anti-aligned in-plane in the AFM phase to aligned out-of-plane in the FM phase with increasing B (black arrows in Fig. 2e).[13,16] Since PC does not respond to B-field, the exciton energy shift leads to a change in the cavity to exciton detuning from 6 meV at 0T to 20 meV at $|B| >2$ T. At the same time, we observe a shift of the LP energy from $1.299\pm0.002$ meV at 0T to $1.306\pm0.002$ meV at $|B| >2$ T, which can be excellently reproduced by Eqs. 1&2 (yellow dotted lines in Fig. 2e) with the corresponding change in detuning ($\Delta=E_{cav}-E_{exc}$) but a constant, keeping the same exciton-cavity coupling strength $g_0$. The same B-dependence of LP energy is confirmed by PL measurements (Fig. 2f). Since the coupling strength sensitively depends on the exciton wavefunction, the unchanged $g_0$ shows that, interestingly, the exciton wavefunction changes negligibly regardless the orientation of the spin anisotropy, as long as the spin anisotropy is maintained.

We furthermore study the temperature-dependence of excitons and polaritons to probe their dependence on the magnetic order. Fig. 2g shows that the normalized reflection spectra at $k_x=0$ as a function of temperature. Both the exciton and polariton modes are clearly resolved with only a small amount of redshift and broadening with increasing temperature up to ~100 K, as expected from bandgap reduction and thermal broadening. Between 100 K and 150 K, however, both the exciton and polariton modes rapidly broaden, and both disappear above 150 K, corresponding to the Néel temperature ($T_N$) of few-layer CrSBr.[16] The gradual redshift and line-broadening below $T_N$, and rapid disappearance of the derivative feature of the exciton above $T_N$, can be more clearly observed in the RC spectra in Supplementary Fig. 6a. This is surprising because 150 K corresponds to a thermal energy of $k_B T \sim 12$ meV, for $k_B$ the Boltzmann constant, which is still much smaller than the exciton binding energy or the exciton-photon coupling strength. Below $T = 120$ K, we can extract the coupling strength (g), exciton energy ($E_{exc}$), and linewidth ($w_{exc}$) as a function of temperature (T), as summarized in Fig. 2h and Supplementary Fig. 6b. The exciton linewidth broadens from $\sim 10$ meV at 5 K to $\sim 20$ meV at 120 K, while g(T) decreases by only 7%, from $57.4 \pm 0.5$ to $53.6 \pm 0.5$ meV. The PC changes negligibly over this temperature range and has $w_{cav} \sim 1$ meV (Supplementary Fig. 7). So the strong coupling condition of $g > (w_{exc} + w_{cav})/4$ is still very well satisfied and the system is expected to remain deep in the strong coupling regime. The rapid disappearance of the exciton and polariton modes above $T_N$ therefore shows clearly that the stability and oscillator strength of the excitons are strongly correlated with the spin anisotropy in CrSBr.

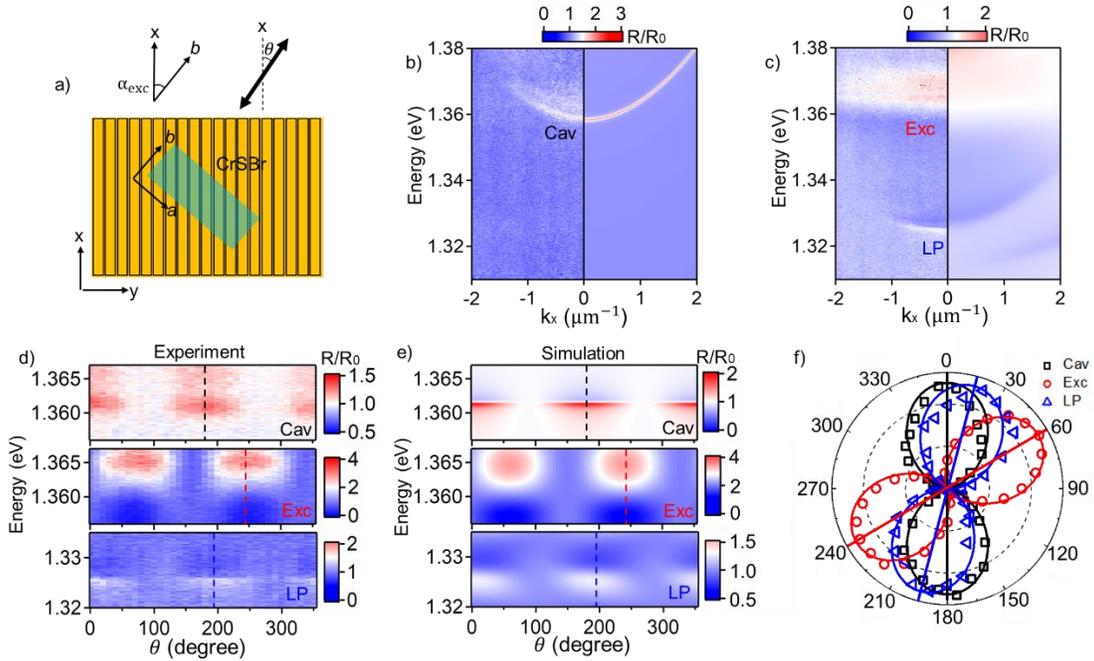

**Fig. 3. Rotation of exciton-polariton polarization due to strongly coupled atomic and photonic anisotropies.** (a) Schematic of 7L-CrSBr flake integrated with PC with a twist-angle (between b-axis of CrSBr and x-axis of PC) of $\alpha_{exc} \sim 60°$. The polarization angle is labeled as $\theta$. Momentum-resolved normalized reflection spectra of (b) bare PC and (c) twisted CrSBr-PC with a twist angle of 60° and polarization angle ($\theta$) of 0°. The left panels are experimental results; right panels are simulation. (d) Polarization angle dependent reflection spectra at $k_x=0$ of the bare cavity (Cav, upper panel), exciton (Exc, middle panel), and LP (lower panel) signals. (e) The simulation of the same data in (d). (f) Polarization angle resolved normalized reflection amplitude of bare cavity, exciton, and LP (symbols).

See Methods for details of calculating the amplitude. The solid curves are simulation. The straight lines mark the fitted polarization direction. See Methods for details of data processing.

The above studies focus on the effect of lattice and spin anisotropies of the crystal. Next we study the effect of strong coupling between photonic crystal lattice and atomic lattice anisotropy by introducing a twist angle $\alpha_{exc} = 60 \pm 3°$ between the b-axis of 7L CrSBr and x-axis of PC (Fig. 3a). Figs. 3b&c show the measured (left panels) and simulated (right panels) momentum-resolved normalized reflection spectra without and with the twisted CrSBr placed on the PC, respectively. Strong coupling and LP band are evident with the CrSBr. Fitting the data with the polariton dispersion (Eq. 1) yields a coupling strength $g(\alpha_{exc}) = 35 \pm 3$ meV $\sim g_0 \cdot \cos(\alpha_{exc})$, corresponding to the dot product of the linearly polarized exciton dipole and cavity field amplitude.

An interesting question is what the polarization of the polaritons is. Figs. 3d and 3e show the measured and simulated polarization-resolved normalized reflection spectra at $k_x=0$ for bare cavity (upper panel), exciton (middle panel), and LP (lower panel), respectively, plotted as a function of $\theta$, the angle of a linear polarization analyzer relative to the x-axis of the PC. Fig. 3f shows a polar-plot of the normalized $\theta$-dependent amplitude of the normalized reflection for the bare cavity, exciton, and LP resonances. All three modes show high degrees of linear polarization. While the cavity and exciton are polarized along their x-axis and b-axis, respectively, the polarization of LP falls in between the two, deviating from the cavity mode by $18 \pm 3°$, and from the exciton mode by $42 \pm 3°$. The measured polarization dependence of the reflectance spectra is well reproduced by RCWA simulation (Fig. 3e). The polarization rotation of the LP modes is further corroborated by simulated distribution of the amplitude and phase of x- and y-components of the LP mode electric field (Supplementary Fig. 9).

The rotation of the LP polarization from those of the exciton and cavity is also reproduced in two 45°-twisted CrSBr-PC devices, both of which show the same LP polarization angle, rotated $8° \pm 3°$ from the cavity, and in agreement with RCWA simulation (Supplementary Figs. 10&11).

Conventionally, anisotropic media inside an isotropic cavity renders emission with the same polarization as the media,[23,28] and vice versa, an isotropic media in an anisotropic cavity renders emission following the cavity anisotropy.[19,20,29] In our twisted CrSBr-PC system, strong coupling correlates the anisotropies of the CrSBr exciton with that of the PC cavity mode, resulting in a polarization rotated from both.

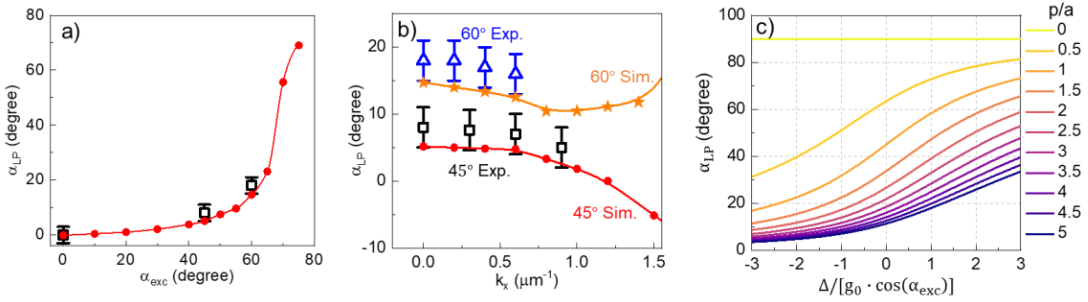

**Fig. 4. Tunable exciton-polariton polarization angle.** (a) Twist-angle dependent LP polarization angle $\alpha_{LP}$ at near zero detuning and zero in-plane wavenumber. The black squares are experimental results; the red filled circles connected with lines are simulation. (b) Momentum dependent LP polarization angle $\alpha_{LP}$ of the 45°- and 60°-twisted CrSBr-PC. The black squares (blue triangles) are experimental results of 45°-twisted (60°-twisted) sample; the red filled circles (orange filled stars) are simulation results of 45°-twisted (60°-twisted) sample, with lines that are guides for the eye. (c) Calculated LP polarization angle $\alpha_{LP}$ as a function of normalized detuning for different p/a, with $\beta=0$, $\alpha_{exc}=60°$, and $g_0$ negative using Eq. S8.

The rotation of the LP polarization is tunable by many degrees of freedom, such as twist-angle, momentum, detuning, and cavity quality factor. Fig. 4a shows the twist-angle dependent LP polarization angle for zero in-plane wavenumber and zero detuning. The simulated $\alpha_{LP}$ agree well with the measured values for devices with both 45° and 60° twists. Fig. 4b shows how $\alpha_{LP}$ changes with momentum ($k_x$) for the 60°- (shown in Fig. 3) and 45°-twisted (shown in Supplementary Fig. 10) CrSBr-PC samples.

To obtain an intuitive understanding of the rotation of LP polarization, we model the system using a 4×4 matrix, including exciton, cavity, the coupling between them, and their coupling to the free space:

$$M = \begin{bmatrix} E_{exc} & g_0 \cdot \cos(\alpha_{exc}) & \beta a \cdot \cos(\alpha_{exc}) & a \cdot \sin(\alpha_{exc}) \\ g_0 \cdot \cos(\alpha_{exc}) & E_{cav} & p & 0 \\ \beta a \cdot \cos(\alpha_{exc}) & p & E_{free,x} & 0 \\ a \cdot \sin(\alpha_{exc}) & 0 & 0 & E_{free,y} \end{bmatrix} \quad (3)$$

Here we set the cavity polarization to be along the x-direction and the exciton polarization is twisted from the x-direction by an angle $\alpha_{exc}$; $E_{free,x}$ and $E_{free,y}$ represent effective free space modes polarized in the x- and y-direction, respectively; $g_0$ is the exciton-cavity coupling strength. a, and p are coupling strengths between exciton and the free space modes, and between the cavity and free space modes, respectively; they are both weak and can be treated perturbatively. The coefficient $\beta \in [0,1]$ counts for the reduced coupling of excitons with the free space mode when the excitons are also coupled to the cavity mode. The larger the exciton-cavity coupling, the closer is $\beta$ to 0.

Diagonalizing only the upper left 2×2 sub-matrix, we obtain LP and UP energies given by the new matrix elements $M_{11}$ and $M_{22}$, respectively, while the sub-matrix of the free-space modes remains diagonal (see Eq. S6 in Methods). The new off-diagonal terms $M_{13}$ and $M_{14}$ ($M_{23}$ and $M_{24}$) now correspond to the coupling of LP (UP) with the free space modes polarized along x- and y-directions, respectively. The polarization angle of the LP (UP) emission is then given by $\arctan(M_{14}/M_{13})$ ($\arctan(M_{24}/M_{23})$). Details of the diagonalization and the expressions of the polarization angles are shown in Methods. .

The resulting $\alpha_{LP}$ differ from the exciton or cavity polarizations and depend on the detuning $\Delta$, $g_0$, p/a, and $\beta$ (Eq. S8), as shown in Fig. 4c. At $\alpha_{exc}$=60° and $\Delta \sim 0$, $\alpha_{LP}$ can be rotated from $\sim 15°$ to $\sim 90°$ as p/a is decreased, such as by increasing the cavity Q. Increasing the Q of our current cavity by two-fold would rotate $\alpha_{LP}$ from ~15° to ~33°. For a fixed p/a, $\alpha_{LP}$ can also be tuned by $\Delta$, such as by tuning the cavity resonance or by tuning the exciton resonance via magnetic field or temperature. For instance, for p/a~0.5, changing the detuning between $\Delta = -g_0 \cdot \cos(\alpha_{exc})$ and $g_0 \cdot \cos(\alpha_{exc})$ rotates $\alpha_{LP}$ by ~22°.

The large tuning range of the polarization angle in our device and the many readily accessible degrees of freedom for tuning are both unusual. Polarization, as a fundamental property of light, plays an essential role in science and technology. However, the polarization of light sources or photonics is difficult to rotate without resorting to bulky optics or strong external fields. The most common way to rotate the light polarization is through the electro-optical[30] or magneto-optical effects.[31] Both effects only induce a polarization rotation angle in sub-degree scale unless a bulky optical medium is used and a high external voltage or field is applied. In contrast, our work shows the possibility to achieve tens of degrees of polarization rotation angle in a few-nm-thick flake integrated on chip. The PC is also only ~100 nm thick and can be readily integrated with other on-chip components.

In short, we demonstrate a twisted light-matter coupled system with strongly correlated spin, atomic, and photonic anisotropies and high turnability. The anisotropy coupling-induced polarization rotation provides a new mechanism for polarization engineering, and may enable compact, on-chip polarized light source[32,33] and polariton lasers.[22,34] The system also provides an example of strongly coupled atomic-scale and photonic-scale orders, which can be extended to other quantum materials or PCs with specific band topologies[8,9] and broken symmetries[35] for photonic engineering of quantum materials.

## Methods

### 1D PC fabrication

The 1D PCs for this work were made by low-pressure chemical vapor deposition following the previously reported procedures.[19,20] The silicon nitride layer is partially etched to form a 1D grating by electron beam lithography followed by plasma dry etching. The grating period $\Lambda$ is 611 nm, the total silicon nitride grating thickness t is 113 nm, the $SiO_2$ capping layer thickness d is 1475 nm, grating step height h is 70 nm, and the gap between grating bars is 50 nm.

### CrSBr single crystal growth

The CrSBr single crystal used in this work was grown using the direct solid−vapor method through a box furnace following the procedures reported previously.[17] Cr powders (Alfa Aesar, 99.97%), S powders (Alfa Aesar,99.5%), and solid $Br_2$ (solidified from liquid bromine, 99.8%, with assistance of liquid nitrogen) were loaded in the clean quartz ampoules inside an Ar-glovebox with the mole ratio of 1:1.1:1.2. The ampoules were subsequently vacuum sealed using the liquid nitrogen trap. The ampoules were heated up to 930 °C very slowly, then held at this temperature for 20 h, followed by cooling down to 750 °C at a rate of 1°C/h. The ampoule was then quenched down to room temperature. Large CrSBr single crystals grew naturally at the bottom of the ampoule and were easily separated from byproducts of $CrBr_3$ that formed at the top of the quartz ampoule.

**CrSBr exfoliation and stacking on PC**

CrSBr flakes were exfoliated using scotch tape. The 7L-CrSBr flakes used in this work were confirmed by atomic force microscopy for the right thickness and transferred to 1D PC using polymer film (1.2g poly(bisphenol A carbonate) dispersed in 20g chloroform) inside a nitrogen-glovebox.

**Optical measurements**

RC and PL spectroscopies are conducted by real-space and Fourier-space imaging of the sample/device. For measurements without magnetic fields, the sample was kept at 5 K unless specified otherwise using a Montana Fusion system. An objective lens with numerical aperture (NA) of 0.42 is used for both focusing and collection. For magnetic field dependent measurements, the sample was kept at 3.5 K using an AttoDry 1000 system, where an objective lens with NA of 0.90 is used for both focusing and collection. A tungsten halogen lamp with a beam size of ~10 μm in diameter is used as the white-light source for all RC measurements. A continuous-wave solid-state laser at 532 nm with a power of 30 μW and a beam size of ~2.5 μm in diameter is used to excite CrSBr samples for PL measurements shown in Figs. 1h&i and Fig. 2f. The density dependent PL measurements (Figs. 2a-d) were carried out using a Tsunami pulsed laser (80 MHz, 5 ps pulse width, 740 nm) with a beam diameter of ~2.5 μm for excitation. The collected signals are all detected by a Princeton Instruments spectrometer with a cooled charge-coupled camera. Polarization dependent RC and PL spectra were collected by rotating the polarization orientation of a linearly polarized white light with a half waveplate and probing the reflection or PL intensity as a function of the rotation angle ($\theta$) with respect to x-axis of PC (scheme in Fig. 3a).

**Electron-hole density estimation**

We made a series of CrSBr flakes with different number of layers (Supplementary Fig. 4) on the top of SiO$_2$/Si substrate and measured their RC spectra (Supplementary Fig. 5a). Then we estimate the absorbance (A) at 1.676 eV (740 nm), the excitation energy for density dependent experiments, from their RC signals according to the reported method:[36]

$$RC = \frac{4}{n_{sub}^2 - 1} A \quad (S1)$$

where $n_{sub}$ is the refractive index of SiO$_2$ at 740 nm (1.45).[37] Then we fit the absorbance of CrSBr with different layer numbers linearly and extract the absorbance for 7L-CrSBr by the fit (Supplementary Fig. 5b). The e-h density per pulse ($n_{eh}$) is then determined by:

$$n_{eh} = [1 - (1-A)^2] \frac{I}{S \cdot h\nu \cdot f} \quad (S2)$$

where I is the excitation power, S is the beam area (~4.9 μm$^2$), h$\nu$ is the excitation photon energy (1.676 eV, 740 nm), f is the repetition rate of the excitation laser pulse (80 MHz). $[1 - (1-A)^2]$ term accounts for the double absorption relation for a reflection measurement.[38] Note that the absorption cross-section reduces at high density,[39] which was not considered in our estimation. Therefore, our extracted $n_{eh}$ and $n_s$ are both the upper limits of their real values.

**Simulation**

The reflection spectra as a function of momentum and polarization are simulated using rigorous coupled wave analysis (RCWA).[40] The electric field distributions of bare cavity and polariton are simulated by Ansys Lumerical FDTD software.[41] The resonances identified by RCWA and FDTD agree with each other. RCWA is much faster for simulating the full spectra as measured in experiments, while FDTD is necessary for analyzing the field profiles of specific resonances of the system. In the simulation, the refraction indices of Si$_3$N$_4$ and SiO$_2$ and Si are set as 2, 1.46, and 3.7. The CrSBr material is modelled as a dielectric material of finite thickness of 6.5 nm whose permittivity tensor is given by:

$$\bar{\bar{\varepsilon}} = \begin{pmatrix} \varepsilon_{bx} + \varepsilon_L & 0 & 0 \\ 0 & \varepsilon_{by} & 0 \\ 0 & 0 & \varepsilon_{bz} \end{pmatrix} \quad (S3)$$

where $\varepsilon_{bx}$, $\varepsilon_{by}$, $\varepsilon_{bz}$ accounts for the dielectric responses from the background other than exciton, while $\varepsilon_L$ denotes the dielectric response from exciton. The anisotropic response is manifested by the anisotropic permittivity tensor, in which only b-axis (x) exhibits dielectric response from exciton. Here, $\varepsilon_L$ takes the forms of a Lorentz oscillator:

$$\varepsilon_L = \frac{f}{E_{exc}^2 - E^2 - iEw_{exc}} \quad (S4)$$

where f is the oscillator strength, $E_{exc}$ denotes the exciton energy, $w_{exc}$ is the FWHM linewidth. To reproduce the results observed in experiment, we set oscillator strength f as 3.0 eV², exciton energy $E_{exc}$ as 1.36 eV, FWHM linewidth ($w_{exc}$) as 10 meV, $\varepsilon_{bx}$, $\varepsilon_{by}$, $\varepsilon_{bz}$ as 2. To introduce a twist-angle ($\alpha_{exc}$) between the b-axis of CrSBr and x-axis of PC, a unitary matrix transformation U is applied, and the permittivity tensor of the twisted CrSBr is shown below:

$$\bar{\bar{\varepsilon}}_\theta = U\bar{\bar{\varepsilon}}U^\dagger,$$

$$U = \begin{bmatrix} \cos(\alpha_{exc}) & -\sin(\alpha_{exc}) & 0 \\ \sin(\alpha_{exc}) & \cos(\alpha_{exc}) & 0 \\ 0 & 0 & 1 \end{bmatrix} \quad (S5)$$

**Polarization-dependent LP signal**

To obtain the polarization-dependent reflectance the polaritons, we first measure the reflectance R of a linearly-polarized beam at the normal incidence with a polarization angle $\theta$ (measured from the cavity polarization direction along $x$-axis), then normalize it by dividing it with the reflectance $R_0$ of the same beam from flat $Si_3N_4$ surface next to the PC to obtain $R/R_0$, and finally subtract the linear background produced by the Fabry-Pérot (FP) effect of the layers. The $R/R_0$ spectra as a function of $\theta$ before subtracting the background subtraction is shown in Supplementary Figs. 8a&b (black solid lines), where the LP resonance leads to a derivative feature above a background that can be fitted with a linear function (red lines in Supplementary Fig. 8a) or a 2$^{nd}$ order polynomial (red lines in Supplementary Fig. 8b). Subtracting the background curve from $R/R_0$ gives the pure LP signal, shown as 2D false color plot of the LP spectra vs. $\theta$ in Fig. 3d and Supplementary Figs. 8d&e. From Supplementary Figs. 8d&e, we confirm the background subtraction is robust regardless of the linear or 2$^{nd}$ order polynomial backgrounds. Comparisons with simulation results are shown in Supplementary Figs. 8c&f and Fig. 3e, showing excellent agreement.

From the LP reflectance spectra obtained above, the amplitude $I_{LP}$ of the LP resonance is given by the difference between the maximum and minimum intensity of the derivative feature of the LP resonance, as illustrated in Supplementary Figs. 8d&e. The amplitude $I_{LP}$ vs. $\theta$ is shown in Fig. 3f of the main text.

The same procedure is used to obtain the polarization properties of the LP resonances of the two 45°-twist CrSBr-PC devices shown in Supplementary Figs. 10-11.

**Polariton polarization model**

To reproduce the measured LP polarization angle with respect to the cavity photon polarization direction (x-axis), we need to quantify the x- and y-component of measured LP intensity after it couples with free space mode. This can be extracted from the 4×4 matrix shown in the main text (Eq. 3), which considers light-matter coupling ($g_0 \cdot \cos(\alpha_{exc})$), exciton-free space mode coupling (a), and cavity photon-free space mode coupling (p). Partially diagonalizing upper left 2×2 matrix of Eq. 3 yields:

$$M' = \begin{bmatrix} E_{LP} & 0 & LP_x & LP_y \\ 0 & E_{UP} & UP_x & UP_y \\ LP_x & UP_x & E_x & 0 \\ LP_y & UP_y & 0 & E_y \end{bmatrix} \quad (S6)$$

with new off-diagonal terms, $LP_x$ ($UP_x$), $LP_y$ ($UP_y$), proportional to the x- and y-components of LP (UP), respectively:

$$LP_x = \frac{-2g_0 \cdot \cos(\alpha_{exc}) \cdot p + \beta a \cdot \cos(\alpha_{exc}) \cdot \left(E_{cav} + \sqrt{4g_0^2 \cos^2(\alpha_{exc}) + (E_{cav} - E_{exc})^2} - E_{exc}\right)}{\sqrt{4g_0^2 \cos^2(\alpha_{exc}) + \left(E_{cav} + \sqrt{4g_0^2 \cos^2(\alpha_{exc}) + (E_{cav} - E_{exc})^2} - E_{exc}\right)^2}},$$

$$LP_y = \frac{a \cdot \sin(\alpha_{exc}) \cdot \left(E_{cav} + \sqrt{4g_0^2 \cos^2(\alpha_{exc}) + (E_{cav} - E_{exc})^2} - E_{exc}\right)}{\sqrt{4g_0^2 \cos^2(\alpha_{exc}) + \left(E_{cav} + \sqrt{4g_0^2 \cos^2(\alpha_{exc}) + (E_{cav} - E_{exc})^2} - E_{exc}\right)^2}},$$

$$UP_x = \frac{-2g_0 \cdot \cos(\alpha_{exc}) \cdot p + \beta a \cdot \cos(\alpha_{exc}) \cdot E_{cav} - \beta a \cdot \cos(\alpha_{exc}) \cdot \left(\sqrt{4g_0^2 \cos^2(\alpha_{exc}) + (E_{cav} - E_{exc})^2} + E_{exc}\right)}{\sqrt{4g_0^2 \cos^2(\alpha_{exc}) + \left(-E_{cav} + \sqrt{4g_0^2 \cos^2(\alpha_{exc}) + (E_{cav} - E_{exc})^2} + E_{exc}\right)^2}},$$

$$UP_y = \frac{a \cdot \sin(\alpha_{exc}) \cdot \left(E_{cav} - \sqrt{4g_0^2 \cos^2(\alpha_{exc}) + (E_{cav} - E_{exc})^2} - E_{exc}\right)}{\sqrt{4g_0^2 \cos^2(\alpha_{exc}) + \left(-E_{cav} + \sqrt{4g_0^2 \cos^2(\alpha_{exc}) + (E_{cav} - E_{exc})^2} + E_{exc}\right)^2}} \quad (S7)$$

Based on Eq. S7 with detuning, $\Delta = E_{cav} - E_{exc}$, we quantify the LP polarization angle ($\alpha_{LP}$) with respect to cavity mode polarization direction:

$$\alpha_{LP} = \arctan\left(\frac{LP_y}{LP_x}\right)$$

$$= \arctan\left(\frac{\Delta + \sqrt{4g_0^2\cos^2(\alpha_{exc}) + \Delta^2}}{-2g_0 \cdot \cot(\alpha_{exc}) \cdot \frac{p}{a} + \beta \cdot \cot(\alpha_{exc}) \cdot \left(\Delta + \sqrt{4g_0^2\cos^2(\alpha_{exc}) + \Delta^2}\right)}\right) \quad (S8)$$

For linearly polarized exciton, cavity photon, and LP modes, a, p, and $g_0$ are all real values, and the relative signs among them, corresponding to the relative phases of 0 or π between the linearly polarized modes, depend on the details of the cavity system.

**Data Availability.** All data that support the results in this article are available from the corresponding author upon request.

**Acknowledgement.** Q.L., A.A. C.B. and H.D. acknowledge the support by the Army Research Office under Awards W911NF−17-1-0312, X.X. Y.L. J.H. and H.D. acknowledge the support by the Air Force Office of Scientific Research under Awards FA2386-21-1-4066, and the National Science Foundation under Awards DMR 2132470. Q.L., X.X. L.Z. K. S. and H.D. acknowledge the support by the Office of Naval Research under Awards N00014-21-1-2770, and the Gordon and Betty Moore Foundation under Grant GBMF10694. W.L. and B.L. acknowledge support by US Air Force Office of Scientific Research (AFOSR) Grant No. FA9550-19-1-0037, National Science Foundation (NSF)- DMREF- 1921581 and Office of Naval Research (ONR) Grant N00014-23-1-2020. This work was performed in part at the University of Michigan Lurie Nanofabrication Facility.

**Author contributions.** Q.L. and H.D. conceived the research. Q.L. conducted the measurements with assistance from J.H.. X.X. fabricated the PCs and conducted the simulations. A.A., Q.L., C.W.B., N.M., and Y.L. exfoliated CrSBr flakes and fabricated CrSBr-PC devices. W.L. L.Z and B.L. provided the CrSBr single crystal. K.S. assisted with theoretical analysis. H.D., K.S., L.Z., and B.L. supervised the project. Q.L., X.X. and H.D. wrote the manuscript. All authors read and commented on the manuscript.

**Competing interests.** The authors declare no competing financial interest.